\documentstyle[12pt]{article}
\begin{document}
\begin{flushright}
CU-TP-598
\end{flushright}
\vspace*{1cm}
\begin{center}
{\Large\bf Color Diffusion and Conductivity}\\
{\Large\bf in a  Quark-Gluon Plasma
}\footnote{ This work was supported by the Director, Office of Energy
Research, Division of Nuclear Physics of the Office of High Energy and Nuclear
Physics of the U.S. Department of Energy under Contract
No DE-FG-02-93ER-40764.}
\\[1cm]
{Alexei V. Selikhov}\footnote{Supported by the AUI/RHIC
Fellowship Fund;\\
Permanent address: {\sl Kurchatov Institute,
123182 Moscow D--182, Russia}}
and {Miklos Gyulassy}\\[2ex]
{\sl Department of Physics, Columbia University, New York, NY10027}\\[2cm]
{ June 1993}\\[1cm]
\end{center}
\begin{abstract}
Color diffusion is shown to be an important  dissipative
property of quark-gluon plasmas that rapidly damps collective
color modes.
We derive the characteristic
color relaxation time scale,
  $t_c\approx (3 \alpha_s T \log(m_E/m_M ))^{-1}$,
showing its  sensitivity  to the ratio of the
static color electric and magnetic screening masses.
This leads to a surprisingly small
color conductivity, $\sigma_c\approx 2 T/\log(m_E/m_M)$,
which in fact vanishes in the semi-classical (1-loop) limit.\\[2ex]
\end{abstract}


The quark-gluon plasma (QGP) phase of QCD matter has been predicted
to exhibit collective behavior similar to  QED plasmas
with Debye screened (color) electric interactions,
collective plasmon oscillations, and collective color mode instabilities
under a variety of non-equilibrium configurations
(see, e.g., reviews in \cite{ElzHei,mrow,qm91}). Such collective behavior
of quark-gluon plasmas (QGP) may  be observed experimentally
via ultra-relativistic
heavy-ion reactions
if the time scales of dissipative processes that act to damp
collectivity are sufficiently long.
Thusfar, most studies of transport properties of QGP
have focused on momentum relaxation processes
to estimate viscosity and  thermal conductivity coefficients, $dE/dx$, etc.
\cite{visc,baym,thoma}.
Relaxation phenomena associated with the color
degrees of freedom have received, on the other hand,  less attention.
In this letter, we derive a new  transport coefficient,
called color diffusion, that has no abelian analog.
The related color relaxation time, $t_c$, measures  how rapidly
the color of a parton precesses in an ambient
fluctuating background field.

The relevance of $t_c$ can be seen most clearly in the context
of color conductivity. In the relaxation time approximation,
classical non-abelian transport theory
leads to the following formula
for the static color
conductivity \cite{Hei86,mrow}
\begin{equation} \sigma_{\rm
c} =  \omega^2_{pl}/\nu
 \; \; , \label{cond} \end{equation}
where $\omega^2_{\rm
pl}=4\pi \alpha_{\rm s} T^2(1+N_f/6)/3$
the color plasmon frequency\cite{Wel,pisar}
and $\nu$ is a collision frequency.
Color conductivity arises
as a generalization of  Ohm's law:
$j^{a}_i=\sigma_c E^a_{i}$
and influences the evolution of a QGP  through ohmic heating
as  pointed out in ref. \cite{gatof}. Ohmic heating
also couples mini-jets and beam jets  in ultra-relativistic
 nuclear collisions as shown in ref.\cite{GyuEsc}.
The main problem in this connection is to estimate the
collision frequency,
$\nu$. In QED plasmas, $\nu=1/t_p$, where
 $t_p$ is the {\em momentum} relaxation time.
In QCD, transport theory estimates for $t_p$ including color
Debye and dynamical screening give\cite{baym}
\begin{equation}
 t_p
\approx (4\alpha_{\rm s}^2 \ln(1/\alpha_{\rm
s})T)^{-1}
 \; \; . \label{taup} \end{equation}
This time scale was used to estimate the magnitude of effects of
ohmic heating in ref.\cite{GyuEsc}.
We show below, however, that $t_c$ instead of $t_p$ controls the color
conductivity,
and that $t_c\ll t_p$ in the perturbative limit.
In fact, we find that $t_c$ is closely related to the (divergent) gluon damping
rate and that color diffusion strongly damps collective
plasmon modes. We therefore find that
non-abelian plasmas are  surprisingly poor color conductors
and that  collective color modes are over-damped due to rapid color diffusion.

To derive $t_c$ we follow  here ref.\cite{Won,Hei85} and
consider the motion of a
classical colored particle moving along a world line  $x^\mu(\tau)$
with momentum $p^\mu(\tau)= m v^\mu(\tau)$
and  $v^\mu=dx^\mu/d\tau$,
that couples with an $N^2-1$ component color charge vector
$\Lambda^a(\tau)$ to a
background  Yang-Mills field, $A^a_\mu(x)$.
We follow this approach here because of its intuitive simplicity
and because it leads to the same result that we find  starting
from the kinetic equations for the quark and gluon Wigner density
matrices~\cite{ElzHei}
including the Lenard-Balescu type collision terms derived in \cite{Sel1,Sel2}.
While the derivation based on the later approach
is more rigorous, it is also less transparent physically and therefore deferred
to a subsequent paper\cite{SelGyu}.

Consider the motion of a heavy quark
in the $d$ dimensional representation
of $SU(N)$ with second order Casimir, $C_2$.
The classical color charge vector is defined through  the expectation
value of the color current via
\begin{eqnarray}
j^{\mu a}(x)=g\langle \bar{\psi}(x)\gamma^{\mu}t^{a}\psi (x)\rangle
= g \int d\tau v^{\mu}(\tau)\delta^{4}(x-x(\tau))
\Lambda^{a}(\tau) \;,
\label{1.2}\end{eqnarray}
where $[t^a,t^b]=if^{abc}t^c$ and $t^a t^a= C_2 {\bf 1}$.
Similarly, the classical momentum is related to the expectation
value of the matter energy-momentum tensor
\begin{eqnarray}
T^{\mu \nu}(x)=i
\langle \bar{\psi}\gamma^{\mu}
D^\nu(x)
\psi (x)\rangle
= m \int d\tau v^{\mu}(\tau)v^\nu(\tau) \delta^{4}(x-x(\tau))
\;,
\label{1.2b}\end{eqnarray}
where $D^\nu(x)=\partial^\nu + ig t^a A^\nu_a(x)$
The equations of motion follow from
the conservation laws
\begin{eqnarray}
\partial_{\mu}T^{\mu\nu} = gj^a_\mu F^{\nu\mu}_a \; \; , \; \;
\partial_{\mu}j^{\mu a}&=&gf^{abc}A_{\mu}^{b}j^{\mu c} \;\; .
\label{tmunu}\end{eqnarray}
Substituting  Eqs.~(\ref{1.2},\ref{1.2b})
into the left and right hand side, we recover
the equations derived in \cite{Won}:
\begin{eqnarray}
\frac{dp^{\mu}}{d\tau}&=&
gv_{\nu}F_{a}^{\mu\nu}\Lambda^{a} \; \; , \label{pmu}\\
\frac{d\Lambda^{a}}{d\tau}&=&
gf^{abc}v^{\mu}A^{b}_{\mu}\Lambda^{c} \;,
\label{1.6}\end{eqnarray}
where the fields are evaluated at $x^\mu(\tau)$.
Since under a gauge transformation
$U(x)=\exp(i\theta^a(x) t^a)$,
$\psi\rightarrow U\psi$, the color charge must transform as
$t^a \Lambda^a(\tau)\rightarrow
U(x(\tau))t^a\Lambda^a(\tau) U^{-1} ( x(\tau))$,
to insure the gauge covariance of eqs.(\ref{pmu},\ref{1.6}).
(Note that $m d/d\tau= p_\mu \partial^\mu$ along the world line.)

We study  the time dependence of the
color charge averaging over an ensemble of background field configurations.
We assume that the ensemble characterizes
a   color neutral medium in which the $A^a_\mu$
are random  fields such that the ensemble averaged potentials
vanish in a suitable gauge.
The ensemble average of Eq.~(\ref{1.6}) gives
\begin{eqnarray}
\frac{d}{d\tau}\langle \Lambda^{a}(\tau)\rangle =
gf^{abc}\langle  [vA]^b_\tau
\Lambda^{c}(\tau)\rangle \;\;,
\label{1.8}\end{eqnarray}
where for shorthand we write $[vA]^b_\tau=(p^{\mu}(\tau)/m)
A^{b}_{\mu}(x(\tau))$.
Integrating (formally) Eq.~(\ref{1.6}) and  substituting
the result  back into Eq.~(\ref{1.8}), yields
\begin{eqnarray}
\frac{d}{d\tau}\langle \Lambda^{a}(\tau)\rangle &=&
g^{2}f^{abc}f^{cde} \int_{0}^{\tau} d\tau^\prime
\langle [pA]^{b}_\tau [p A]^{d}_{\tau^\prime}
\Lambda^{e}(\tau^\prime)\rangle \label{1.9a}\end{eqnarray}
Our first main physical assumption
is that the time scale of field fluctuations
is short compared to the time
scale of variations of the color orientation.
Formally, this is suggested   from eq.(\ref{1.9a}) {\em if} $g\ll 1$
 because the rate of change
of the color  charge is  proportional to $g^2$.
We therefore assume
 a stochastic (random phase) ansatz\cite{Kam}
for the expectation value of the product
\begin{eqnarray}
\langle [vA]^{b}_\tau [vA]^{d}_{\tau^\prime}
\Lambda^{e}(\tau^\prime)\rangle \approx
\langle [vA]^{b}_\tau [vA]^{d}_{\tau^\prime}\rangle
\langle\Lambda^{e}(\tau)\rangle \; \;.
\label{1.9b}\end{eqnarray}
With this ansatz, eq.(\ref{1.9a}) reduces to
\begin{eqnarray}
\frac{d}{d\tau}\langle \Lambda^{a}(\tau)\rangle &\approx& \left\{g^{2}
f^{abc}f^{cde} \int_{0}^{\tau} d\tau^\prime
\langle [vA]^{b}_\tau[vA]^{d}_{\tau^\prime}\rangle
 \right\}
\langle \Lambda^{e}(\tau)\rangle \equiv -d^{ae}(\tau)
\langle \Lambda^{e}(\tau)\rangle \; \; .
\label{1.9}\end{eqnarray}
This defines the color diffusion tensor, $d^{ae}(\tau)$,
whose physical interpretation will become clear below.
Note that we cannot use the
$v^\mu A^{b}_\mu =0$ gauge in Eq.~
(\ref{1.9}) because
 $v^{\mu}(\tau)$, via eq.(\ref{pmu}),
 is different for every
 member of the ensemble.

Our second main approximation
is to work in the eikonal limit.
This is equivalent to the ``hard thermal loop'' approximation
in \cite{pisar}. We thus
assume that our test parton has a  high initial
four momentum
and that energy loss via eq.(\ref{pmu}) is small.
This also follows formally from  eq.(\ref{pmu}) in the perturbative $g\ll 1$
limit.
In this case, we can factor the four velocity out of the ensemble average
and approximate
\begin{eqnarray}
\langle [vA]^{b}_\tau [vA]^{d}_{\tau^\prime}\rangle
\approx v^\mu v^\nu
\langle A^{b}_\mu(x(\tau)) A^{d}_\nu(x(\tau^\prime)) \rangle
\equiv v^\mu v^\nu C^{bd}_{\mu\nu}(x(\tau),x(\tau^\prime))
\; \; . \label{dab}\end{eqnarray}
Next we note that in this eikonal limit $x^\mu(\tau)
-x^\mu(\tau^\prime)=(\tau-\tau^\prime) v^\mu$,
and thus for a  homogeneous,
color neutral ensemble, the correlation function $C$ can expressed as
\begin{eqnarray}
C^{bd}_{\mu\nu}(x(\tau),x(\tau^\prime))
=\delta_{bd} \int (dk) e^{-i(kv)(\tau-\tau^\prime)}
C_{\mu\nu}(k,u)
\; \; , \label{dbdk} \end{eqnarray}
where $(dk)=d^4 k/(2\pi)^4$,
$u^\mu$ is the four velocity of the ensemble rest frame.
The correlation function $C$ measures the spontaneous
fluctuations of the background field.
In the classical limit those fluctuations can be calculated
using kinetic theory\cite{Hei86,Sel2,Sil71}.
For a system in thermal equilibrium,
the $\omega \ll T$ spontaneous fluctuations can also be related to
 the response function (retarded commutator)
via the fluctuation-dissipation theorem\cite{callen} as
\begin{eqnarray}
C^{\mu\nu}(k,u)\approx -\frac{2T}{ku}
 {\rm Im} D_R^{\mu\nu}(k,u,T)
\; \; ,
\label{fd} \end{eqnarray}
where
$$D_R^{\mu\nu}(k,u,T)= i \int d^4 x e^{ikx} \theta(x_0)
{\rm Tr}\left(\rho [A^\mu(x),A^\nu(0)]\right)$$
 is the thermal averaged retarded commutator that
arises in linear response theory. The statistical operator is
given by $\rho=\exp(-P^\mu u_\mu/T)$
for a thermal ensemble with four velocity $u$.
Both methods give the same result
in the high temperature limit in the 1-loop approximation\cite{Hei86}.
The 1-loop result obtained in ref.\cite{Wel,pisar} is
\begin{eqnarray}
D_{R\mu\nu}(k,u)=
- \frac{Q_{\mu\nu}}{k^2-\Pi_L}
- \frac{P_{\mu\nu}}{k^2-\Pi_T}  + \alpha \frac{k_\mu k_\nu}{k^2}
\; \; , \label{dk} \end{eqnarray}
where $\alpha$ is a gauge parameter.
This decomposition utilizes the longitudinal
and transverse projectors,
$Q_{\mu\nu}=\bar{u}_\mu\bar{u}_\nu/\bar{u}^2$
and $P_{\mu\nu}=\bar{g}_{\mu\nu}-Q_{\mu\nu}$, as given  in terms
of $\bar{g}_{\mu\nu}=g_{\mu\nu}-k_\mu k_\nu/k^2$
and $\bar{u}_\mu=\bar{g}_{\mu\nu}u^\nu$.
Furthermore, the longitudinal and transverse
polarization functions are related to the gluon self energy
$\Pi^{\mu\nu}(k)$ through
$\Pi_L=\Pi^{\mu\nu}Q_{\mu\nu}$ and $\Pi_T=\Pi^{\mu\nu}P_{\mu\nu}/2$.

With (\ref{fd},\ref{dk}), the color diffusion tensor
can be expressed as
\begin{eqnarray}
d^{ae}(\tau)= -2g^2 N T \delta_{ae} \int \frac{(dk)}{(ku)} {\rm Im}
(v D_R v)
\int_0^\tau d\tau^\prime e^{-i(kv)(\tau-\tau^\prime)}
\; \; , \label{dae2} \end{eqnarray}
where the color factor for $SU(N)$ follows from  $f^{abc}f^{ebc}=N
\delta_{ae}$.
As $\tau\rightarrow \infty$, the proper time
integral reduces to $\pi\delta(kv)$ as can be seen  using
${\rm Im} D_R(-k)=-{\rm Im} D_R(k)$,
to replace the limits of integration
by $-\tau$ to $\tau$.
We can therefore define the color diffusion coefficient, $d_c$,
and the corresponding color relaxation time in the plasma rest frame,
$t_c$, via
\begin{eqnarray}
\lim_{\tau\rightarrow \infty} d^{ae}(\tau)\equiv \delta_{ae} N d_c
\equiv \delta_{ae} (v u)/ t_c\; \; .
\label{dc} \end{eqnarray}
For ultra-relativistic partons, the color relaxation rate reduces to
\begin{eqnarray}
\frac{1}{t_c}&=&-2\pi g^2 N T
\int \frac{(dk)}{(ku)} \frac{k^2}{(ku)^2-k^2} \delta\left(\frac{kv}{vu}\right)
{\rm Im} \left( \frac{1}{k^2-\Pi_L} -
\frac{1}{k^2-\Pi_T} \right)
\; \; , \label{dc2} \end{eqnarray}
It is important to note that the dependence on the gauge parameter
$\alpha$ dropped out  because the delta function
constrains $kv=0$.
However, in the semi-classical limit $(\omega\ll T)$
the gauge invariance of $t_c$ holds even if $kv\ne 0$, as can be shown
via kinetic theory\cite{SelGyu}, since fluctuations are driven by
gauge invariant current-current correlations (polarization tensor).
The gauge invariance of this result is related
to the gauge invariance of the damping rate in
hard thermal loop calculations in ref.\cite{pisar2}
and depends essentially on the eikonal limit.
For momenta $\sim 3T$, the eikonal approximation
applies perturbatively because  the
energy loss per interaction, $\sim g^2 T$, is relatively small
if $g\ll 1$.

Given $t_{c}$, the solution of eq.~(\ref{1.9}) in the plasma rest frame is
\begin{eqnarray}
\langle\Lambda^{a}(t)\rangle =
e^{-t/t_c} \langle\Lambda^{a}(0)\rangle \; .
\label{1.19}\end{eqnarray}
Note, that while the  ensemble averaged color of the
parton vanishes rapidly for $t>t_c$,
the equations of motion conserve the magnitude
of the color vector, i.e., $\langle\Lambda^{a}(\tau)\Lambda^{a}(\tau)\rangle =$
constant.

We can compare the color relaxation rate, $1/t_c$,
to the quark or gluon damping rates
noting\cite{Wel} that the longitudinal and transverse polarization functions
are related to the color dielectric functions via
$k^2-\Pi_L=k^2\epsilon_L$ and
$k^2-\Pi_T=\omega^2 \epsilon_T - \vec{k}^2$.
The analytic expressions for $\epsilon_L(k)$ and $\epsilon_T(k)$
are recorded for example in \cite{ElzHei,mrow}.
In terms of these color dielectric functions
\begin{eqnarray}
\frac{1}{t_c}&=&-\frac{\alpha_s N T}{2\pi^2}
\int \frac{d^3k}{\omega k^2}
{\rm Im} \left( \frac{1}{\epsilon_L(k) } +
\frac{k^2-\omega^2}{\omega^2 \epsilon_T(k) - k^2} \right)_{\omega=
k\cos\theta}
\; \; . \label{dc3} \end{eqnarray}
This is the same integral that occurs
in the quark damping rate, $\gamma_q$, as discussed in ref.\cite{thoma}.
In fact we find that it corresponds exactly
to the leading log approximation to the gluon damping rate
computed in \cite{pisar2}.
This coincidence of results based on classical kinetic theory
and 1-loop high T QCD is a general result\cite{Hei86}
in all problems  where  the classical $\omega\ll T$ and $k\ll T$
modes of the system are dominant.

Unfortunately, as is well known\cite{pisar2}, the transverse
contribution to that damping rate is logarithmically divergent.
Dynamical screening is sufficient to regulate the
Coulomb divergence in the expression for the momentum relaxation
rate\cite{baym} but not the damping rate.
The longitudinal part is finite\cite{thoma,pisar2}
and can thus be neglected. Several attempts to deal with this
problem have been proposed. One is to introduce a nonperturbative
magnetic mass, $m_M\sim g^2 T$, as in ref.\cite{linde}.
Another is to introduce damping self consistently by adding
an imaginary part to the bare fermion propagator\cite{altherr}.
However, no satisfactory solution is yet available\cite{peigne}.

In our kinetic formulation the source of the problem
may be traced back to our first assumption that the field fluctuation  time
is short compared to the color diffusion time. What we learned a postiori
was that  fluctuations of the transverse magnetic fields
are long ranged in time and that quasi-static unscreened magnetic fields lead
to the divergent color relaxation rate.
This suggests that the change of the color moment with
time should taken into account within the integrand
in eq.(\ref{1.9}).
Such a ``memory'' effect could damp
the contribution from early times with $(vu)(\tau-\tau^\prime) > t_c$.
Inserting a factor
$\exp(-(vu)(\tau-\tau^\prime)/t_c)$ within the integral over
$\tau^\prime$ would smear out the delta function,
$\delta(kv)$, appearing in eq.(\ref{dc2}) into  a Lorentzian
form $\propto d_c/((kv)^2+ d_c^2)$.
The resulting  self consistent equation for $d_c$ would lead to
a  finite result analogous to the method considered
by Altherr et al.\cite{altherr}.

We will however follow Pisarski\cite{pisar2}
and regulate the divergence by
introducing a magnetic mass cutoff.
In the region $\omega\ll k$, we therefore approximate
\begin{eqnarray}
{\rm Im} \left(
\frac{1}{\omega^2 \epsilon_T(k) - k^2} \right) \approx
- \frac{a\omega/k}{(k^2+m_M^2)^2 + (a\omega/k)^2}
\; \; , \label{mag}\end{eqnarray}
where $a=3\pi \omega_{pl}^2/4\approx m_E^2$.
In the $m_M=0$ limit this coincides with the one-loop form
used for example in \cite{thoma}.
Note that the dynamic screening factor\cite{baym} is properly
included above. With eq.(\ref{mag}) we can integrate
the transverse part to obtain
\begin{eqnarray}
\frac{1}{t_c}&\approx& \frac{\alpha_s N T}{\pi}
\int \frac{dkk^2 d\mu}{\mu k^3}
 \frac{a\mu k^2}{(k^2+m_M^2)^2 + (a\mu)^2}
\approx
  \alpha_s N T \log(m_E/m_M)
\; \; , \label{mag2}\end{eqnarray}
where the final form is  numerically accurate already for $m_M<m_E$.
The finite longitudinal part adds about 0.5 to the log
which is well within the uncertainties associated with inserting
a magnetic mass in (\ref{mag}) by hand.

To show that $d_c$ actually corresponds to a diffusion coefficient
in kinetic theory, we consider next
 the  classical phase space density $Q(x,p,\Lambda)$ of
an ensemble of colored particles defined by
\begin{eqnarray}
Q(x,p,\Lambda)= \int d\tau' \sum_{i} \delta (x-x_{i}(\tau'))
\delta (p-p_{i}(\tau')) \delta (\Lambda-\Lambda_{i}(\tau')) \;\; ,
\label{klim}\end{eqnarray}
where  $p^\mu_i$ and $\Lambda^a_i$
obey  Wong equations (\ref{1.6}) coupled via a self consistent field.
The phase space density obeys the (Heinz) transport equation
\cite{ElzHei,Hei85}:
\begin{eqnarray}
(p_\mu \partial_x^{\mu}+g p^\nu F_{\mu\nu}^a\Lambda^a \partial_p^{\mu}
+i g p^\mu A^a_\mu L^a )Q(x,p,\Lambda) = 0 \;\; ,
\label{kineq1}\end{eqnarray}
where
\begin{eqnarray}
L^{a}= -\imath f^{abc}\Lambda^{b} \frac{\partial}{\partial \Lambda^{c}}  \; ,
\label{1.2.5}\end{eqnarray}
are generators of $SU(N)/Z^N$ obeying  the commutation relations
$[L^{a}, L^{b}]=\imath f^{abc}L^{c}$.
eq Note that (\ref{kineq1}) just  non-abelian Louiville equation
expressing that the parton density is constant
along a characteristic, ($x(\tau),p(\tau),\Lambda(\tau))$,
satisfying the Wong equation in the self  consistent field,
i.e.,
\begin{eqnarray}
\frac{d}{d\tau}Q(x(\tau),p(\tau),\Lambda(\tau))=
-\int d\tau' \frac{d}{d\tau'} \sum_i \delta(x(\tau)-x_i(\tau'))
\cdots =0 \;\;.
\label{1.2.1}\end{eqnarray}

The ensemble average of eq.~(\ref{1.2.1}) can be expressed as
\begin{eqnarray}
m \frac{d}{d\tau}\langle Q(x(\tau),p(\tau),\Lambda(\tau)) \rangle=
-\imath g \langle p^{\mu}\delta A^{a}_{\mu}
L^{a} \delta Q \rangle - g \langle p^{\nu} \delta F^{a}_{\mu\nu}
\Lambda^{a} \partial^{\mu}_{p}\delta Q\rangle \;\;,
\label{kineq3}\end{eqnarray}
where we decompose $Q=\langle Q \rangle+\delta Q$
and $A=\langle A \rangle + \delta A$,
and the characteristic trajectory satisfies
\begin{eqnarray}
m \frac{dx^\mu}{d\tau}= p^\mu \;\;, \;\;
\frac{dp^\mu}{d\tau}= g v_\nu \langle F^{\mu\nu}_a \rangle \Lambda^a \;\;,\;\;
\frac{d\Lambda^a}{d\tau}=g f^{abc} v^{\mu}
\langle A^b_{\mu}\rangle \Lambda^c \;.
\label{avwong}\end{eqnarray}
To proceed, we assume average color neutrality,
 $\langle A\rangle=\langle F\rangle=0$, and neglect
the momentum degradation part of the kinetic equation to focus
exclusively on  color diffusion. Therefore, the mean phase space
density evolves according to
\begin{eqnarray}
m\frac{d}{d\tau}\langle Q(x(\tau),p,\Lambda) \rangle=
-\imath g \langle v^{\mu}\delta A^{a}_{\mu}
L^{a} \delta Q \rangle \;\; .
\label{1.2.7}\end{eqnarray}
The kinetic equation for
the fluctuating part of the phase space density
is obtained by subtracting  eq.(\ref{kineq3}) from .(\ref{1.2.1}):
\begin{eqnarray}
\frac{d}{d\tau} \delta Q(x(\tau),p(\tau),\Lambda(\tau))=
-\imath g [v \delta A ]^{a}_{\tau}
L^{a} \langle Q \rangle - g  v^{\nu} \delta F^{a}_{\mu\nu}
\Lambda^{a} \partial^{\mu}_{p} \langle Q\rangle \;\;,
\label{kineq4}\end{eqnarray}
 where in this case
\begin{eqnarray}
\frac{dp^\mu}{d\tau}= g v_\nu (\langle F^{\mu\nu}_a \rangle
+\delta F^{\mu\nu}_a) \Lambda^a \;\;,\;\;
\frac{d\Lambda^a}{d\tau}=g f^{abc} p^{\mu} (\langle A^b_{\mu}\rangle
+\delta A^b_{\mu}) \Lambda^c \;.
\label{flwong}\end{eqnarray}
Taking into account our assumption $\langle A \rangle =
\langle F \rangle =0$, neglecting the induced fluctuations caused
by the gradient in momentum, and retaining
only terms linear in  fluctuating quantities
Eq. (\ref{kineq4}) reduces to
\begin{eqnarray}
\frac{d}{d\tau} \delta Q(x(\tau),p,\Lambda)=
-\imath g [v \delta A ]^{a}_{\tau}
L^{a} \langle Q \rangle \;\; ,
\label{1.2.9}\end{eqnarray}
This linearized approximation
implicitely assumes the eikonal limit since then $p$ is constant
and also $\Lambda$ is constant to that order.
Thus  both the assumptions
made in deriving the color diffusion coefficient
previously correspond to this linearization of the kinetic equations
for fluctuations.
Integrating eq.(\ref{1.2.9}) and substituting the
induced fluctuation into eq.(\ref{1.2.7}) leads to
\begin{eqnarray}
\frac{d \langle Q \rangle}{d\tau} = - g^2 \int_{-\infty}^{\tau}
d\tau' \langle [v \delta A]^a_{\tau}[v \delta A]^b_{\tau'} \rangle L^a L^b
\langle Q(x(\tau'),p,\Lambda) \rangle \;\; .
\label{1.2.9a}\end{eqnarray}

Assuming that fluctuation correlation time is short compared to the
color diffusion time,
$\langle Q(x(\tau'),p,\Lambda)\rangle \approx
\langle Q(x(\tau),p,\Lambda) \rangle$ allowing us to
factor $Q$  out of the integrand. In this way,
we obtain  for  homogeneous and color neutral
ensemble the following  color diffusion equation
\begin{eqnarray}
\frac{d}{d\tau}\langle Q(x(\tau),p,\Lambda) \rangle=
-{d}_{c}L^{a}L^{a} \langle Q \rangle \;,
\label{1.2.10}\end{eqnarray}
where color diffusion coefficient ${d}_{c}$ is given by
\begin{eqnarray}
{d}_c = g^2 v^{\mu}v^{\nu} \int^0_{-\infty} d\tau'
C_{\mu\nu}(x(0),x(\tau'))  \;\;,
\label{1.2.10a}\end{eqnarray}
where $C$ is given by eqs.(\ref{dab}-\ref{dk}),
is thus the same color diffusion coefficient that we derived
via  eqs.(\ref{dae2}-
\ref{mag2})).
Note that
Eq.~(\ref{1.2.10}) can be
written in the form of Fokker-Planck equation
\begin{eqnarray}
\frac{d\langle Q \rangle}{d\tau}=- \frac{\partial}{\partial\Lambda^{a}}
(d^{ab} \frac{\partial}{\partial\Lambda^{b}}\langle Q \rangle) \;,
\label{1.2.11}\end{eqnarray}
where color diffusion coefficient $d^{ab}$ is given by
\begin{eqnarray}
d^{ab}= {d}_{c}f^{adc}f^{ceb}\Lambda^{d}\Lambda^{e} \; .
\label{1.2.12}\end{eqnarray}
Equation~(\ref{1.2.10}) therefore describes diffusion in color space.
Note that operator
$C_{2}=L^{a}L^{a}$ just
is quadratic Casimir operator of $SU(N)/Z^N$.
For a more detailed derivation of collision terms
including momentum space diffusion we refer to \cite{Sel1,Sel2,SelGyu}.

To illustrate color diffusion, it is instructive to consider $SU(2)$,
where $L^{i}=-\imath\epsilon^{ijk}\Lambda^{j}\partial/
{\partial\Lambda^{k}}$
is an angular momentum operator in color (iso-spin) space and
$C_{2}=l(l+1)$.
Since the magnitude of $\vec{\Lambda}$ is fixed,
any initial distribution in color space can be expanded
in the spherical harmonics as
\begin{eqnarray}
\langle Q (\vec{\Lambda},\tau=0)=
\sum_{l,m} c_{l,m} Y^m_l(\theta,\phi) \; \; , \label{1.2.17}\end{eqnarray}
where $\theta,\phi$ are polar coordinates in iso-spin space.
In this case, the proper time evolution is simply
\begin{eqnarray}
\langle Q(\vec{\Lambda},\tau) \rangle=\sum_{l,m}  c_{l,m} Y^m_l(\theta,\phi)
e^{-{d}_{c}\tau l(l+1)} \; \; .
\label{1.2.18}\end{eqnarray}
{}From Eq.~(\ref{1.2.18}) it follows that
any non-isotropic distribution in color space
evolves to a uniform distribution,
\begin{eqnarray}
\lim_{\tau \rightarrow \infty}\langle Q(\vec{\Lambda},\tau) \rangle
=c_0/\surd{4\pi} \;,
\label{1.2.22}\end{eqnarray}
on a proper time scale $\sim 1/2d_{c}$ (for $N=2$).
The same picture holds qualitatively for the general $SU(N)$ case,
where any  distribution evolves to uniform distribution which is
determined by the zero mode of quadratic Casimir operator. This color
equilibrium state is achieved after a proper $\tau_{c}=1/Nd_{c}$.
Since the time in the plasma rest frame is $t=(vu)\tau$ and $Nd_c
=(vu)/t_c$, color equilibrium in the plasma rest frame
is achieved on the time scale $t_c$.

Finally, to calculate the color conductivity coefficient,
we must test the linear response of the system to
a weak external field. In an external filed, $F_{ex}$,
the kinetic equation for the mean density must be supplemented
by a Vlasov term, $gp F_{ex} \Lambda\partial_p\langle Q\rangle$,
on the left hand side of Eqs.(\ref{1.2.7},\ref{1.2.10}).
To calculate the induced color current, we
multiply that modified Eq.(\ref{1.2.10})
with $\Lambda^a$ and integrate over the color charge with
the measure~\cite{Hei85}
\begin{eqnarray}
d\Lambda \propto \prod_{i=1}^{N^2-1}
d\Lambda_{i} \delta (\Lambda^a \Lambda_a -q_{2})
\delta (\Lambda^{a}\Lambda^{b}\Lambda^{c}d_{abc}-q_{3}) \;.
\label{1.2.25}\end{eqnarray}
The proportionality constant is chosen such that~\cite{Hei85}
$\int d\Lambda=1$,
$\int d\Lambda \Lambda^a=0$, and $\int d\Lambda \Lambda^a \Lambda^b
= c_2 \delta^{ab}$. The charges
are defined through the trace of the Casimir operators:
$q_2={\rm Tr}(t^a t^a)$, $q_3=d_{abc} {\rm Tr}(t^a t^b t^c)$.
For quarks $q_2=(N^2-1)/2$,
$q_3=(N^2-4)(N^2-1)/4N$ and $c_2=1/2$. For gluons $q_2=N(N^2-1), q_3=0, c_2=N$.

This leads to the following linearized
kinetic equation for the adjoint color moment density, $q^a(x,p)$,
\begin{eqnarray}
v^{\mu}\partial_{\mu} q^a +
c_{2} gv^{\mu} [F_{ex}]^a_{\nu\mu}  \partial^{\nu}_p
\langle Q \rangle_0 = -N {d}_{c} q^{a} \; .
\label{kinvlas}\end{eqnarray}
where
\begin{eqnarray}
q^{a}(p,x)=\int d\Lambda \Lambda^{a} \langle Q(x,p,\Lambda) \rangle \;\;,
\label{1.2.24}\end{eqnarray}
For a homogeneous, color neutral ensemble, $\langle Q(p,x,\Lambda) \rangle=
\langle Q(p) \rangle_0$, and thus since  $\int d\Lambda \Lambda^a=0$
the color moment density vanishes. However, in a weak external
field there is a distortion of the color density linear in $F_{ex}$.
The factor $N$ on the right hand side of (\ref{kinvlas})
follows from  $\int d\Lambda \Lambda^a L^2 Q=N q^a$,
which can be verified  using
$[L^{a},\Lambda^{b}]=\imath f^{abc} \Lambda^{c}$
and the property that $\int d\Lambda L^a F(\Lambda)=0$,
which follows upon integration by parts
and noting the symmetry properties of $f^{abc}$ and $d^{abc}$.
We note also that in contrast to (\ref{kinvlas})
the kinetic equation for the color singlet
density, $q(x,p)=\int d\Lambda \langle Q\rangle$, is  independent
of the color diffusion coefficient.

Solving (\ref{kinvlas}) for the Fourier transform, $q^a(k,p)$,
 the induced color current is given by
\begin{eqnarray}
j^{\mu a}(k)=g\int dp p^{\mu} q^a(p,k)
\equiv \sigma^{\mu\alpha\beta}(k) [F_{ex}(k)]^a_{\alpha\beta} \;\; ,
\label{defcond}\end{eqnarray}
where the conductivity tensor
for a homogeneous color neutral plasma of quarks, antiquarks, and gluons
is~\cite{Hei86,mrow}
\begin{eqnarray}
\sigma^{\mu\alpha\beta}(k) = g^2\int dp \frac{p^{\mu}
p^{\alpha}}{-i(pk) + (p u)/t_c} \partial^{\beta}_p \{\frac{N_f}{2}(f_q(p)
+f_{\bar{q}}(p)) + N f_g(p)\}
 \;\;,
\label{conten1}\end{eqnarray}
where, $f_i(p)$ are the momentum space distributions including
spin degeneracy for the partons. For example, in equilibrium
$f_g(p)= (2/(2\pi)^3) 2 \theta(p_0)\delta(p^2)/(e^{pu/T}-1)$.
For an isotropic plasma this tensor reduces to
$\sigma^{\mu\alpha\beta}=\sigma^{\mu\nu}(k) u^{\beta}$.
Finally, from Eq.(\ref{conten1}) it follows that~\cite{Hei86,mrow}
the static $(k=0)$ color conductivity is $\sigma^{\mu\nu}(0)=\sigma_c
g^{\mu\nu}$,  where
\begin{eqnarray}
\sigma_c=t_c \omega^2_{pl} \;\;.
\label{concoef}\end{eqnarray}
The important difference between our derivation and that
in ref.\cite{Hei86,mrow} is that the collision rate is
derived  here directly from the kinetic equation
for color fluctuations rather than parameterized
via the relaxation time approximation.
We see in particular that the color diffusion time
rather than momentum degradation time
controls the magnitude of the conductivity.
In addition, the same color diffusion rate applies for both quarks and
gluons unlike in the general ansatz of \cite{mrow}.
Inserting the expression for $t_c$, we thus find the
surprising result stated in the introduction, i.e., that
\begin{equation}
\sigma_c\approx  2T/\log(m_E/m_M)
\; \;.
\end{equation}
The appearance of the non-perturbative and non-classical
magnetic mass in this final expression suggests
that the classical $(k\rightarrow 0)$
or high $T$ limit of a QGP is in fact be very different
from QED plasmas because of color diffusion.
Another indication of this is that the
long wavelength $(k\ll T)$ color plasmon mode, obtained from the dispersion
relation $k^2=\Pi_L(k)=-iQ_{\mu\nu}
(\sigma^{\mu\alpha\nu}-\sigma^{\mu\nu\alpha})k_\alpha$,
are strongly damped with a rate proportional to $1/t_c$
as can be seen from the general discussion in ref.\cite{Hei86,mrow}.
Thus, the anomalously large damping rate
of hard thermal partons propagates down
to the soft collective modes. Probably not only the plasmon
but also the plasmino are over-damped due to color diffusion.
In this respect, further study of color diffusion and screening
of transverse interactions are very important in the
development of QGP transport theory. In practical applications,
especially to nuclear collisions, of course all time integrals
are restricted to a finite time, which automatically
regulates the formal static divergences. The
problem in that connection is to determine which effect, finite time
or  nonperturbative
magnetic screening, restricts the magnitude of color diffusion
the most.\\[2ex]

Acknowledgments: We thank R. Pisarski, S. Gavin, and T.Ludlam  for
fruitful  discussions.\\



\begin{thebibliography}{99}

\bibitem{ElzHei}  H.-Th. Elze and U. Heinz, Phys. Rep. 183 (1989) 81;
U. Heinz, Phys. Rev. Lett., 51 (1983) 351;
H.-Th. Elze, M. Gyulassy and D. Vasak, Nucl. Phys.
B276 (1986) 706; Phys. Lett. 177B (1986) 402.

\bibitem{mrow} S. Mr{\'o}wczynski, in
{\it Quark-Gluon Plasma}, ed. R. Hwa, World Scientific, 1990;
Phys.Rev. D39 (1989) 1940.

\bibitem{qm91} See refs. in
 Proc. of Quark Matter 91, Nucl. Phys. A544 (92) 1c.

\bibitem{visc}
P. Danielewicz and M. Gyulassy,
Phys.\ Rev.\  {\bf D31} (1985) 53;
 A. Hosoya and K. Kajantie,
Nucl.\ Phys.\  {\bf B250} (1985) 666.

\bibitem{baym}
G. Baym, et al.
Phys.\ Rev.\ Lett.\ {\bf 64} (1990) 1867.

\bibitem{thoma} M. Thoma, M. Gyulassy, Nucl. Phys. B351 (1991) 491;
E. Braaten, M. Thoma, PRD 44 (1991) 1298; R2625;
M. Gyulassy, et al. Nucl. Phys. A 538 (1992) 37c.


\bibitem{Hei86} U. Heinz, Ann. Phys. (N.Y.) 168 (1986) 148;


\bibitem{Wel} H.A. Weldon, Phys. Rev., D26 (1982) 1394;
V.V. Klimov, Sov. J. Nucl. Phys. 33 (1981) 934;
E.V. Shuryak, JETP 47 (1978) 212.

\bibitem{pisar} E. Braaten, R. Pisarski, Phys.Rev.Lett. 64 (1990) 1338;
Nucl.Phys. B337 (1990) 569.

\bibitem{gatof} G. Gatoff, A. K. Kerman and T. Matsui,
Phys.\ Rev.\ {\bf D36} (1987) 114.

\bibitem{GyuEsc} K.J. Eskola and M. Gyulassy, Phys.Rev. C47 (1993) 2329.

\bibitem{Won} S.K. Wong, Nuovo Cim. 65A (1970) 689.

\bibitem{Hei85} U. Heinz, Ann. Phys. (N.Y.) 161 (1985) 48.


\bibitem{Sel1} A.V. Selikhov, Phys. Lett., B268 (1991) 263;
               Erratum B285 (1992) 398.


\bibitem{Sel2} A.V. Selikhov, Preprint IAE-5526/1, Moscow,1992;
                  to be submitted to Nucl. Phys. B.

\bibitem{SelGyu} A.V. Selikhov and M. Gyulassy, in preparation.

\bibitem{Kam} N.G. van Kampen, Phys. Rep. C24 (1976) 171.


\bibitem{Sil71} V.P. Silin, Sov.Phys. J.E.T.P. 11
(1960) 1136.


\bibitem{callen} H.B. Callen and T. A. Welton, Phys.Rev. 83 (1951) 34;
W.B. Thompson and J. Hubbard, Rev.Mod.Phys. 32 (1960) 714;
J. Hubbard Proc. Roy. Soc. A260 (1961) 114.


\bibitem{pisar2} R.D. Pisarski, Phys.Rev.Lett. 63 (1989) 1129;
R.D. Pisarski, BNL-P-1/92, unpublished.

\bibitem{linde} A. Linde, Phys. Lett. 93B (1980) 327;
D. J. Gross, R.D. Pisarski, L.G.Yaffe, Rev.Mod.Phys. 53 (1981) 43.

\bibitem{altherr} V.V. Lebedev and A.V. Smilga, Ann.Phys. (N.Y.)
202 (1990) 229;
T. Altherr, E. Petitgirard, T. del Rio Gaztellurutia,
Phys. Rev. D47 (1993) 703;
R. Kobes, G. Kunstatter, K. Mak, PRD45 (1990) 4632;

\bibitem{peigne} S. Peigne, E. Pilon, D. Schiff, LPTHE-Orsay-93/13
unpublished.

\end{thebibliography}
\end{document}